\documentstyle[multicol,aps,prb,epsf]{revtex}
\begin{document}

\draft

\title{
A crib-shaped triplet pairing gap function 
for an orthogonal pair of quasi-one 
dimensional Fermi surfaces in Sr$_2$RuO$_4$
}

\author{
Kazuhiko Kuroki,\cite{UEC} Masao Ogata, Ryotaro Arita,
and Hideo Aoki 
}

\address{Department of Physics, University of Tokyo, Hongo,
Tokyo 113-0033, Japan}

\date{\today}

\maketitle

\begin{abstract}
The competition between spin-triplet and singlet pairings is 
studied theoretically for the tight-binding 
$\alpha$-$\beta$ bands in Sr$_2$RuO$_4$, which arise from 
two sets of quasi-one dimensional Fermi surfaces. 
Using multiband FLEX approximation, where 
we incorporate an anisotropy in the spin fluctuations
as suggested from experiments, we show that 
(i) the triplet can dominate over 
the singlet (which turns out to be extended s), and 
(ii) the triplet gap function optimized in the Eliashberg 
equation has an unusual, very non-sinusoidal form, 
whose time-reversal-broken combination 
exhibits a crib-shaped amplitude with dips. 
\end{abstract}

\medskip

\pacs{PACS numbers: 74.20-z, 74.20Mn}

\begin{multicols}{2}
\narrowtext

Spin-triplet superconductivity is of great conceptual 
importance. It is interesting to ask how the spins in a Cooper 
pair can align, since usually the singlet pairing is favored so 
that some special mechanism should be envisaged to 
account for a triplet superconductivity.  
The $p$-wave pairing in superfluid $^3$He is an outstanding example, 
where a clear picture of the hard-core interaction favoring 
the triplet exists.  

In the past several years, Sr$_2$RuO$_4$\cite{Maeno} 
has attracted much attention as a strong candidate for 
triplet superconductivity.  
In a seminal paper, Rice and Sigrist suggested a mechanism 
for triplet pairing in this 
material, which they call an `electronic version of He'.\cite{RS} 
In their scenario the orbital degeneracy causes ferromagnetic spin 
fluctuations, which is considered to favor the triplet pairing.  
Subsequent experiments indeed suggested triplet pairing.\cite{Luke,Ishida}
However, a new puzzle arose when 
the spin fluctuation in Sr$_2$RuO$_4$ was found to be
antiferromagnetic rather than ferromagnetic 
in a neutron scattering experiment.\cite{Sidis} 
Usual wisdom dictates that antiferromagnetic
spin fluctuations lead to singlet $d$-wave pairing. \cite{MS}

Recently, Kuwabara and one of the present authors,\cite{KO} 
and independently Sato and Kohmoto,\cite{SK} 
have proposed that anisotropy in the spin-fluctuation, 
observed in NMR experiments for Sr$_2$RuO$_4$,\cite{Mukuda} 
may lead to triplet $p$-wave pairing.
However, simple functional forms for 
triplet and singlet gap functions were
assumed in ref.\onlinecite{KO}, i.e., $\sin k_x(+i\sin k_y)$\cite{MN} 
for $p$-wave and $\cos k_x -\cos k_y$ for  $d_{x^2-y^2}$-wave.
In ref.\onlinecite{SK}, the form of the gap 
when the quasi-1D Fermi surfaces are hybridized 
was discussed only qualitatively.
The form of the gap function is 
crucial in discussing the triplet-singlet competition, 
so the functional form should be optimized.  
The form of the gap function is also important in comparing
with experimental results because, 
e.g., some recent experiments have suggested presence of 
nodes in the superconducting gap,\cite{Nishizaki,Ishida2} 
while a thermal conductivity measurement suggests 
an isotropic gap.\cite{Matsuda}

The purpose of the present paper is to determine the functional form of the 
superconducting gap function self-consistently for the spin-anisotropy 
mechanism mentioned above. We adopt 
the fluctuation exchange (FLEX) approximation\cite{Bickers}, 
in which we incorporate anisotropy in the spin fluctuations. 
The FLEX results are then fed into the linearized Eliashberg equation to
obtain the gap function.   
The optimized triplet gap function turns out to have 
unexpected, non-sinusoidal forms, 
which are in sharp contrast with the $k_x$ or $\sin k_x$
gap functions assumed previously.
As a result, the amplitude of the gap has a shape of a 
crib along the Fermi surface, which 
may resolve the controversial experiments. 
The origin of the peculiar behavior of the gap function is traced back 
to the singular $k$-dependence in the spin susceptibility due to 
a nesting of two sets of quasi-one dimensional Fermi surfaces.  

The ruthenate is essentially a three-band system, 
which arises from $d_{xz}$ orbitals aligned linearly along the $x$ axis, 
$d_{yz}$ along the $y$, and $d_{xy}$ in the $xy$ plane.  
The former two give rise to the $\alpha,\beta$ bands, 
which are well nested due to the quasi-one dimensionality and 
causes antiferromagnetic spin fluctuations.  
In this paper, we concentrate on the $\alpha,\beta$ bands, namely, 
we consider a tight-binding model, 
\begin{eqnarray}
H&=&-t\sum_{\sigma}\sum_{m=xz,yz}
\sum_{\langle ii'\rangle}^{\rm nn}
\left(c^{m\dagger}_{i\sigma}c^{m}_{i'\sigma}+{\rm H.c.}\right)\nonumber\\
&-&t'\sum_{\sigma}\sum_{\langle i,j\rangle}^{\rm nnn}
\left(c^{xz\dagger}_{i\sigma}c^{yz}_{j\sigma}+{\rm H.c.}\right)
+U\sum_i\sum_{m=xz,yz} n^{m}_{i\uparrow}n^{m}_{i\downarrow}.
\end{eqnarray}
on a square lattice.
Here $c^{m\dagger}_{i\sigma}$ creates 
an electron at $d_{m} (m = xz$ or $yz$) orbital.
the nearest-neighbor hopping integral $t$ is along the $x (y)$ direction for 
$d_{xz} (d_{yz})$ orbitals.  
We take $t=1$ as a unit of energy.  

We have also included the 
next nearest-neighbor hopping $t'$ which corresponds to a weak hybridization.
When $t'\neq 0$ the two sets of quasi-one-dimensional 
bands anticross, and two 
two-dimensional (rounded-square) bands result, which are 
the $\alpha$ and $\beta$ bands.  
The on-site repulsive interaction, $U$, is considered within
each orbitals, and interorbital interactions are neglected 
for simplicity.\cite{Takimoto} 
The band filling is $n=4/3$ electrons per orbital in 
Sr$_2$RuO$_4$.  

In treating the interaction, we employ the FLEX approximation.  
This method is a kind of self-consistent random-phase approximation (RPA) 
where the dressed Green's function is used in the RPA diagrams.  
In the multiband version of FLEX,\cite{Koikegami,Kontani}  
the Green' function $G$, 
the susceptibility $\chi$, the self-energy $\Sigma$, and 
the superconducting gap function $\phi$ all become $2\times 2$ matrices,
e.g., $G_{lm}({\bf k}, i\varepsilon_n)$, where $l,m$ denote
$d_{yz}$ or $d_{xz}$ orbitals. 
The orbital-indexed matrices for Green's function and the gap functions
can be converted into band-indexed ones with a unitary transformation. 
As for the spin susceptibility, we diagonalize the 
$2\times 2$ matrix $\chi^{zz}_{\rm sp}$ and 
concentrate on the larger eigenvalue, denoted as $\chi^{zz}$.  

The actual calculation proceeds as follows: 

\noindent (i) Dyson's equation is solved to obtain the 
renormalized Green's function $G(k)$, 
where $k$ is a shorthand for the wave vector ${\bf k}$ and 
the Matsubara frequency, $i\epsilon_n$, 

\noindent (ii) The fluctuation-exchange interaction 
$V^{(1)}(q)$ is given as\cite{Vcom}
\begin{equation}
V^{(1)}(q)=\frac{1}{2}V^{zz}_{\rm sp}(q)+V^{+-}_{\rm sp}(q)+
\frac{1}{2}V_{\rm ch}(q).
\label{v1}
\end{equation}  
The effective interactions due to longitudinal $(zz)$ and 
transverse $(+-)$ spin fluctuations (sp) and that due to charge 
fluctuations (ch) have the forms
$V^{zz}_{\rm sp}=U^2\chi^{zz}_{\rm sp}$,
$V^{+-}_{\rm sp}=U^2\chi^{+-}_{\rm sp}$, and 
$V_{\rm ch}=U^2\chi_{\rm ch}$, respectively, 
where the spin and the charge susceptibilities are 
\begin{eqnarray}
\chi^{zz}_{\rm sp}(q)&=&\chi^{\rm irr}(q)[1-U\chi^{\rm irr}(q)]^{-1}, \\
\chi^{+-}_{\rm sp}(q)&=&
\alpha\chi^{\rm irr}(q)[1-U\alpha \chi^{\rm irr}(q)]^{-1},\\ 
\chi_{\rm ch}(q)&=&\chi^{\rm irr}(q)[1+U\chi^{\rm irr}(q)]^{-1}, 
\label{spchchi}
\end{eqnarray}
in terms of the irreducible susceptibility 
$\chi^{\rm irr}(q)= -\frac{1}{N}\sum_k G(k+q)G(k)$
($N$:number of $k$-point meshes).
Here we have taken account of the anisotropy in the spin fluctuation by 
introducing a phenomelogical parameter $\alpha$ \`a la 
Kuwabara-Ogata.\cite{KO} From the NMR experiments,\cite{Mukuda}
we assume $\alpha <1$.

\noindent (iii) $V^{(1)}$ then brings about the self-energy, 
$\Sigma(k)=\frac{1}{N}\sum_{q} G(k-q)V^{(1)}(q)$,
which is fed back to Dyson's equation, 
and the self-consistent iterations are 
repeated until convergence is attained.
We take $64\times 64$ $k$-point meshes and 
up to 16384 Matsubara frequencies in order to
ensure convergence at low temperatures. 

We determine $T_c$ as the temperature at which the eigenvalue $\lambda$ of 
the {\'E}liashberg equation, 
\begin{eqnarray}
\lambda_\mu\phi_{\mu l m}(k)\nonumber &=& -\frac{T}{N}\sum_{k'}\sum_{l',m'}\\
&&\times V_{\mu l m}^{(2)}(k-k')G_{ll'}(k')G_{mm'}(-k')\phi_{\mu l'm'}(k'),
\label{eliash}
\end{eqnarray}
reaches unity.
Here the pairing interaction $V_\mu^{(2)}$ 
is given by 
\begin{equation}
V_s^{(2)}=\frac{1}{2}V^{zz}_{\rm sp}+V^{+-}_{\rm sp}
-\frac{1}{2}V_{\rm ch}
\label{pairs}
\end{equation}
for singlet pairing, 
\begin{equation}
V_{t\perp}^{(2)}=
-\frac{1}{2}V^{zz}_{\rm sp}-\frac{1}{2}V_{\rm ch}
\label{pairt1}
\end{equation}
for triplet pairing with $S_z=\pm 1$ ($\vec{\it d}\perp\vec{z}$), and 
\begin{equation}
V_{t\parallel}^{(2)}=\frac{1}{2}V^{zz}_{\rm sp}-V^{+-}_{\rm sp}
-\frac{1}{2}V_{\rm ch}
\label{pairt0}
\end{equation}
for triplet pairing with $S_z=0$ ($\vec{\it d}\parallel\vec{z}$).
Here $\vec{\it d}$ is the $d$-vector characterizing 
the triplet pairing gap function.

When the electron-electron repulsion, which causes fluctuations, 
is short-ranged (as for the Hubbard $U$ interaction) 
the spin fluctuations are much stronger than the charge 
fluctuations, i.e., $(V_{\rm sp} \gg V_{\rm ch})$.\cite{Takimoto}
When fluctuations for a certain (`nesting') 
wave vector {\bf Q} are pronounced, 
the main contribution in the summation in eq.(\ref{eliash}) 
comes from those satisfying ${\bf k-k'\simeq Q}$, 
which should be the case when the Fermi surface is nested.  
The present Fermi surface is indeed well nested due to the 
quasi-one-dimensionality. 

\begin{figure}
\begin{center}
\leavevmode\epsfysize=170mm \epsfbox{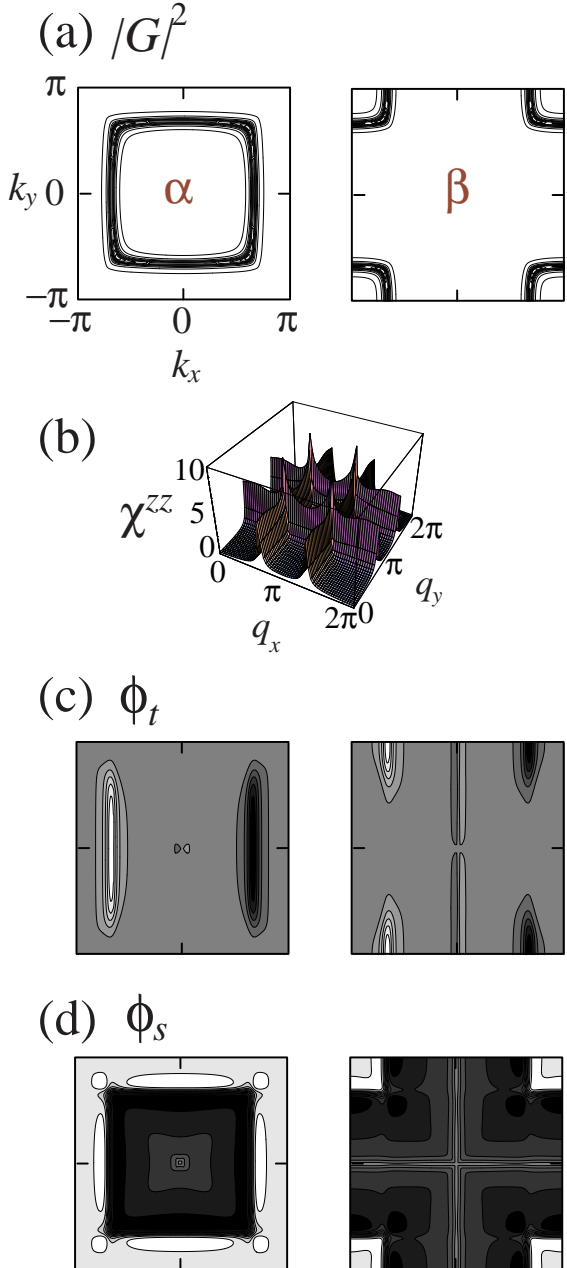}
\caption{The contour plot of the 
FLEX result for Green's function ($|G|^2$; a), 
spin susceptibility (b), optimized gap function 
for the spin triplet ($\phi_t$; c) or singlet ($\phi_s$; d) pairing 
for $U=5$, $t'=0.3$, $\alpha=0.8$, and $T=0.02$.  
The left(right) panel for the $\alpha(\beta)$ band, and 
white(black) corresponds to positive(negative) amplitude in (c,d). 
}
\label{fig1}
\end{center}
\end{figure}

Now we turn to the results summarized in Fig.\ref{fig1} 
for $t'=0.3$ and $U=5$.  
In Fig.\ref{fig1}(a), the ridges in $|G|^2$ 
delineate $\alpha$ and $\beta$ Fermi surfaces.
These quasi-1D surfaces are strongly nested 
at ${\bf q}\equiv (2\pi/3,q_y)$ and ${\bf q}\equiv (q_x,2\pi/3)$ 
(mod $2\pi$), so that the spin
susceptibility is ridged in a crib shape as shown in Fig.\ref{fig1}(b), 
with peaks at the corners, ${\bf q}\equiv(\pm 2\pi/3,\pm 2\pi/3)$.  
This is consistent with neutron scattering experiments.\cite{Sidis}

The triplet and singlet 
gap functions obtained by solving the Eliashberg
equation are shown in Fig.\ref{fig1}(c) and (d), respectively. 
Remarkably, the triplet gap function takes a strange shape: 
although the symmetry is consistent with $p$-wave, its form 
is far from $\sin k_x$ along the rounded-square 
Fermi surface. Rather, it has an almost constant amplitude 
on a pair of parallel sides, $k_x\equiv\pm 2\pi/3$ (mod $2\pi$), 
of the square with a {\it vanishing} amplitude 
on the other pair ($k_y\equiv\pm 2\pi/3$) 
of parallel sides.  This applies to each of the $\alpha$ and 
$\beta$ bands.  Of course the symmetry dictates that 
the other solution ($p_y$), rotated by 90 degrees from what is described 
here ($p_x$), enters on an equal footing as we shall discuss below.  

Why do we have this peculiar behavior for the 
FLEX+Eliashberg optimized gap function ?  
To begin with, superconductivity arises due to pair scattering  
from $({\bf k,-k})$ to $({\bf k',-k'})$ 
mediated by the pairing interaction $V^{(2)}({\bf q})$, 
where ${\bf q=k-k'}$ is the momentum transfer.
From the BCS gap equation we can see that superconductivity 
arises if the quantity 
\begin{equation}
V_{\phi}=-\frac{\sum_{{\bf k,k'}\in {\rm FS}} V^{(2)}_\mu({\bf k-k'})
\phi_\mu({\bf k})\phi_\mu({\bf k'})}
{\sum_{{\bf k}\in {\rm FS}} [\phi({\bf k})]^2}
\label{vpp}
\end{equation}
is positive and large, where we denote the gap function as $\phi_\mu$ 
($\mu=s$ for singlet and $t$ for triplet pairing).
As discussed by Kuwabara and Ogata\cite{KO}, 
and independently by Sato and Kohmoto\cite{SK}, the $p$-wave pairing, 
with $\phi_t({\bf k})\phi_t({\bf k+Q})<0$, is favored 
when spin anisotropy is so strong as to 
realize $V^{zz}_{\rm sp}({\bf Q})>2V^{+-}_{\rm sp}({\bf Q})$, i.e.,
$V_{t\parallel}({\bf Q})>0$.
However, the present self-consistent calculation shows that 
the situation is a little more involved.
A key factor is the spin fluctuation that is 
enhanced along a line

\begin{figure}
\begin{center}
\leavevmode\epsfysize=114mm \epsfbox{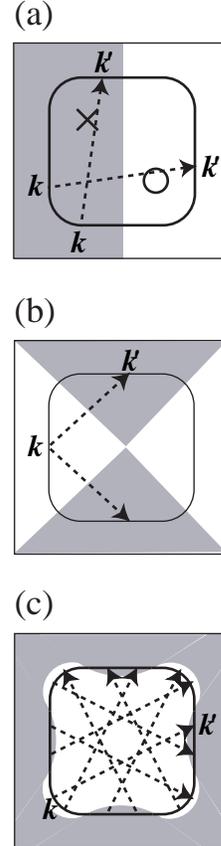}
\caption{Pair scatterings across the Fermi surface 
which favor $(\bigcirc)$ or unfavor $(\times)$
$p_x$-wave pairing (a), those contributing to $d_{x^2-y^2}$ (b) or to 
extended $s$ pairings (c). The white(grey) areas 
represent positive(negative) $\phi$.
}
\label{fig2}
\end{center}
\end{figure}

\noindent
${\bf q}\equiv (q_x,2\pi/3)$.  
For a given nodal line (vertical for $p_x$ pairing), 
the pair scatterings across one pair of parallel sides of the 
Fermi surface become all 
favorable as indicated by $\bigcirc$ in Fig.\ref{fig2}(a).
This gives rise to the near-constant gap function on that pair of sides
of the Fermi surface.  By contrast, 
the pair scatterings across the other pair of parallel 
sides ($\times$ in Fig.\ref{fig2}(a)) lead to 
\[
-V^{(2)}_t(k_x-k'_x,-\frac{2\pi}{3}-\frac{2\pi}{3})
\phi_t(k_x,-\frac{2\pi}{3})\phi_t(k'_x,\frac{2\pi}{3})<0
\]
in eq.(\ref{vpp}) when $k_x,k_x'$ have the same sign.
Since $-V^{(2)}_t$ is negative, the $p_x$-wave gap functions on these
sides of the Fermi surface is unfavored. 
This explains the gap function shown in Fig.\ref{fig1}(c).

Now, let us move on to the competing superconducting state 
in the singlet channel.
Although Kuwabara and Ogata discussed a competition
between the triplet $p_x$-wave and a singlet $d_{x^2-y^2}$-wave,
we find here that the real competitor (the most stable singlet state) 
is unexpectedly
an extended $s$-wave rather than $d_{x^2-y^2}$-wave.
This is understood as follows.
Since $V_s({\bf Q})>0$ is repulsive, 
$\phi_s({\bf k})\phi_s({\bf k+Q})<0$ has to
be satisfied. For the $d_{x^2-y^2}$-wave, 
the ${\bf k,k'=k+Q}$ on the Fermi surface that

\begin{figure}
\begin{center}
\leavevmode\epsfysize=40mm \epsfbox{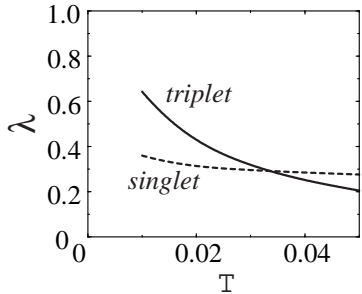}
\caption{Eigenvalues of the Eliashberg equation for 
triplet and singlet pairings as a function of temperature 
for $U=5$, $t'=0.3$, $\alpha=0.8$.}
\label{fig3}
\end{center}
\end{figure}

\begin{figure}
\begin{center}
\leavevmode\epsfysize=50mm \epsfbox{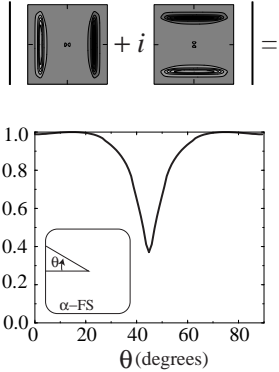}
\caption{$|\phi({\bf k})|$ (eq.(\protect\ref{abs})) 
for $U=5$, $t'=0.3$, $\alpha=0.8$, and $T=0.02$.  
The inset shows how the two functions are combined.}
\label{fig4}
\end{center}
\end{figure}

\noindent
satisfy ${\bf k-k'}\sim (\pm 2\pi/3,\pm 2\pi/3)$ (mod$(2\pi,2\pi)$)
and $\phi_s({\bf k})\phi_s({\bf k'})<0$ are only 
${\bf k, k'}\sim(\pm 2\pi/3,0),(0,\pm 2\pi/3)$ 
(dashed arrows in Fig.\ref{fig2}(b)). 
By contrast, all the ${\bf k,k'}$'s on the Fermi surface with 
${\bf k-k'}\sim (\pm 2\pi/3,\pm 2\pi/3)$ 
contribute to the extended s-wave pairing (Fig.\ref{fig2}(c)).

The competition between the triplet and singlet is 
quantified by the eigenvalues of the Eliashberg equation, 
shown in Fig.\ref{fig3} as  
functions of temperature.\cite{commentLL} It can be seen that for the value of 
$\alpha(=0.8)$ adopted here, the triplet does dominate at low temperatures.
The singlet extended-s, 
although having a greater magnitude of the pairing interaction,
has the nodal line running in the vicinity of the 
Fermi surface (Fig.\ref{fig1}(d)), which should be why this pairing is 
weaker. 
If we extrapolate $\lambda$ to $T\rightarrow 0$ 
a finite transition temperature much smaller than $O(0.01t)$ is suggested.

The above result and argument are for the $p_x$-wave.  
Obviously, the $p_y$-wave should 
be degenerate with it from symmetry.  
A complex linear combination of these two, $p_x+i p_y$, 
which breaks the time reversal symmetry, should be the true state 
below $T_c$ since the superconducting gap is maximized for that 
combination. 

The absolute value of 
the gap function for that linear combination 
on the Fermi surface, 
\begin{equation}
|\phi({\bf k})|=\left[\phi_t(k_x,k_y)^2+\phi_t(k_y,k_x)^2\right]^{1/2},
\label{abs}
\end{equation}
along the Fermi surface is displayed in Fig.\ref{fig4},
which has a crib shape with dips at 
the corners around ${\bf k}\equiv (\pm 2\pi/3,\pm 2\pi/3)$.\cite{dip}
The dip arises because $\phi_t$ in each of the $p_x$ and $p_y$ channels 
has already sharp drops at the corner of the Fermi surface.  
Thus, in phase-insensitive experiments, the gap function obtained here 
may look like a two-dimensional $f$-wave pairing with 
nodes along $k_x=\pm k_y$, since a dip and a node are 
indistinguishable when the temperature is greater 
than the dip. 
Further study on this point is under way.

Discussions with Yuji Matsuda are gratefully acknowledged.  
This work is in part funded by the Grant-in-Aid for Scientific
Research from the Ministry of Education of Japan.

\end{multicols}

\begin{references}
\bibitem{Vcom} In eqs.(\protect\ref{v1}),(\protect\ref{pairs})
$-$(\protect\ref{pairt0}) 
we have omitted the first- and second-order terms, 
$U$ and $-U^2\chi_0$ (in eq.\protect\ref{v1}), 
which are negligible when spin and/or charge fluctuations are strong, 
although we have taken those terms into account in the actual calculation.
\bibitem{commentLL} 
The simultaneous enhancement of the triplet pairing and the spin 
fluctuations is reminiscent of the Tomonaga-Luttinger theory for 
purely 1D systems, where one can show that the
susceptibilities for the triplet pairing and the SDW
may diverge simultaneously for $T\rightarrow 0$ {\em when} the SU(2)
symmetry in the interaction is broken.
However,  in the Tomonaga-Luttinger case, 
the triplet pairing dominates over the SDW only when the
electron-electron interaction is attractive, in sharp contrast with the
present case where the interaction is
repulsive.  The difference comes from a different origin of the broken
SU(2) here, which is the anisotropy in the irreducible susceptibility 
(which arises at a single electron level) 
rather than the anisotropic electron-electron interaction.
\bibitem{dip} We note that, while the qualitative result 
does not depend crucially on $t'(=0.3$ in this paper), the quantitative 
shape of the dip does.
\end{references}
\end{document}